\documentstyle[prb,aps]{revtex}
\begin{document}
\draft

 \title{Spin-flip transitions between Zeeman sublevels  in semiconductor quantum dots}

 \author{Alexander V. Khaetskii \cite{address} and Yuli V. Nazarov}
\address{ Faculty of Applied Sciences and DIMES, Delft University of
Technology,\\
Lorentzweg 1, 2628 CJ Delft, The Netherlands}

\maketitle

\begin{abstract}

We have studied spin-flip  transitions between Zeeman sublevels
in GaAs electron quantum dots.
Several different mechanisms which originate
from spin-orbit coupling are shown to be responsible for such processes.
 It is shown that spin-lattice relaxation for
 the  electron localized in a quantum dot is much less effective than for the free electron. 
The spin-flip rates due to several other mechanisms not related to the spin-orbit interaction 
are also estimated.  

\end{abstract}


\section{Introduction.}

\par
 Quantum dots (QD's)  in semiconductor heterostructures
provide a unique opportunity to study
the properties of the electron quantum states in detail and manipulate
the electrons in these "artificial atoms" in a controllable way (see reviews
\cite{1,2}).
The shape and size
of  quantum dots can be varied by changing the gate voltage.
This also tunes  the number of electrons in the dot. Besides, the
electronic states can be significantly modified by a magnetic field applied
perpendicular to the plane of the heterostructure.
\par
Quantum dots are considered  as possible candidates for
building a quantum computer.\cite{diVincenzo}
The crucial point of the idea is the necessity to couple dots coherently and  keep coherence on sufficiently long time scales.
In this respect, there is a great demand in the
theoretical estimation of the typical spin dephasing time of the electron in the QD.
In our previous work \cite{Khaetskii}
we have shown that
 the localized character of the electron wave functions  in the QD's
suppresses the most effective intrinsic spin-flip
mechanisms related to the absence of  inversion
symmetry in  GaAs-like crystals. This leads to an unusually low rate of  spin-flip transitions.
However, in Ref. \onlinecite{Khaetskii} we  concentrated on the case of inelastic transitions between
the neighbouring quantized energy levels in the dot which corresponds
 to a relatively large energy transfer.
On the other hand, the quantum bit was proposed to involve two Zeeman sublevels of the same orbital level.
Therefore, in the present work we consider the transitions between such sublevels.
Since the transition involves a fairly small energy transfer, the resuls are very
different from those of Ref. \onlinecite{Khaetskii}.
 \par
As in Ref. \onlinecite{Khaetskii}, we concentrate on the  spin-flip processes
due to the spin-orbit interaction.
 This is the main source of  the spin-flips for the three- and two-dimensional electron states
in the  GaAs-type crystal without an inversion center.
Besides, in such a  polar-type crystal
one finds a strong coupling of electrons to the bosonic environment via
the piezo-electric interaction with acoustic phonons.
The combination of these two mechanisms provides an effective
spin-lattice relaxation of {\it free} carriers in $A_{III}B_{V}$ 
semiconductors and heterostructures.\cite{Pikus}
We  show, however, that
the spin-lattice relaxation for the  electron localized
     in the  QD   is much less effective.

We have calculated the rates for the different spin-orbit related mechanisms which cause a spin-flip
in the course of the phonon-assistant transition  between the Zeeman sublevels.
Besides, we have estimated the spin-flip rates due to several other mechanisms, 
for example, due to
the fluctuating magnetic field produced by the  fluctuating electron density 
in  the leads 
or due to 
 modulation of the hyperfine coupling with nuclei by lattice vibrations.

\section{Spin-orbit mechanisms.}

\par
We consider 
the case of a strong confinement in the z-direction  and the typical 
lateral dot  size  is of the order of  thousand $\AA$ and much 
larger than the width of the 2D layer in the z-direction.
We begin with the following one-electron Hamiltonian that is derived
from the Kane model (see Ref. \onlinecite{Pikus})
and describes the  2D electrons of the conduction band
in the presence of magnetic field ${\bf B}$,
   lateral confining potential $U({\bf r})$
and the phonons:
   \begin{eqnarray}
\hat {\cal H}= \frac{\hat{\bf p}^2}{2m}+ U({\bf r})+U_{ph}({\bf r},t)
+\frac{1}{2}g\mu_B {\hat{\mbox{\boldmath $\sigma$}}}\cdot {\bf B} +\sum_{i=1}^{3} \hat {\cal H}_i;
   \nonumber \\
                                   \hat {\cal H}_1=
\beta (-\hat\sigma_x \hat p_x + \hat\sigma_y \hat p_y); \/
\beta=\frac{2}{3} \langle p^2_z \rangle \frac{\Delta}
{(2mE_g)^{1/2} m_{cv} E_g}; \
\hat {\cal H}_2=\frac{1}{2} V_0  {\hat{\mbox{\boldmath $\sigma$}}}
\cdot {\hat {\mbox{\boldmath $\varphi$}}}; \
\hat {\cal H}_3= \tilde g \mu_B \sum_{i\neq k} u_{ik} \hat \sigma_i B_k
 \label{1}
   \end{eqnarray}
Here $\hat {\bf p}=-i\hbar {\mbox{\boldmath $ \nabla$}} +(e/c) {\bf A} $
  is the 2D  electron  momentum operator, $m$  the effective mass,
 $\hat  {\mbox{\boldmath $\sigma$}} $ the Pauli matrices.
Axes $x,y,z$ coincide with the main crystallographic ones with the z-axis
along the normal to the 2D plane (the [100] orientation).
 The magnetic field has an arbitrary direction.
The third term describes the spin-independent interaction with the phonons,
including the piezoelectric ones. The fourth term is
the Zeeman energy.
 The other three terms describe all possible spin-orbit effects.
$\hat {\cal H}_1$ stems from the absence of the
inversion symmetry in the bulk.\cite{Dyak}
Velocity $\beta$
 takes the values in the interval
 $(1\div 3) \cdot 10^5 cm/s$ for GaAs heterostructures.
   $\hat {\cal H}_2$ describes the spin-orbit splitting of the electron spectrum due
   to the strain field
produced by  the acoustic phonons.
There $\hat \varphi_x = (1/2)\{u_{xy},\hat p_y \}_+$,
$\hat \varphi_y = -(1/2)\{u_{yx},\hat p_x \}_+$,
$\hat \varphi_z = (1/2)\{u_{zx},\hat p_x \}_+ - (1/2)\{u_{zy},\hat p_y \}_+$,
where  $\{,\}_+$ denotes the anticommutator, $u_{ij}$
is the lattice strain tensor, and  $V_0 = 8\times 10^7 cm/s$.
\cite{Titkov}
  In  GaAs  the electron $g$-factor ($g=-0.44$) differs  strongly
from the free electron value $g_0=2$ owing to the strong  spin-orbit interaction
which mixes the valence band  and conduction band states \cite{Kittel}.
The admixture  depends on the lattice deformation.
Coefficient $\tilde g$ can be found within the Kane approach,
$ \tilde g = (2 m_0/\sqrt{3}m)
(\Delta/E_g) (d/E_g)$,
$d= -4.5$  eV is one of the three deformation  constants describing
the strain effect on the hole band splitting
\cite{Pikus}, $\tilde g \approx 10.4$.

\par The three terms in Hamiltonian (\ref{1}) correspond to the three distinct
mechanisms of the spin flip.
The first mechanism  is due to the spin-orbit admixture of state with an
opposite spin.
While without the spin-orbit interaction the Zeeman sublevels
correspond to the orbital state with the spin up or down,
the spin-orbit terms provide a small admixture of the state of
the opposite spin to each sublevel.
This enables the phonon-assistant transition
 between the two states. This mechanism corresponds to 
  term $\hat {\cal H}_1$.
\par
The second and the third mechanisms  are described by the $\hat {\cal H}_2$ and
$\hat {\cal H}_3$ terms and correspond to two distinct kinds of  direct spin-phonon coupling.
Below we show that the admixture mechanism is actually a dominant one.

\subsection{Admixture mechanism.} \label{admixture}

Let us show that for this mechanism
the matrix element of the $U_{ph}({\bf r},t)$ operator
for the spin-flip transition between the Zeeman levels is  proportional
to the product of the Zeeman energy and the phonon strain field.
Since we deal with a small energy transfer, we consider only the interaction
with piezo-phonons, hence, for  mode ${\bf q}\alpha$, $\alpha =l,t$ , we have
\cite{Levinson}
\begin{equation} 
U_{ph}^{{\bf q}\alpha}({\bf r},t) = \sqrt{\hbar/2\rho \omega_{{\bf q}\alpha}}
\exp(i{\bf qr}-i \omega_{{\bf q}\alpha}t) eA_{{\bf q}\alpha}b_{{\bf q}\alpha}^+ 
+ c.c.;\/  A_{{\bf q}\alpha}=
 \xi_i\xi_k \beta_{ikj} e^{j}_{{\bf q}\alpha},
 \label{Int}
\end{equation} 
where the effective piezo-electric modulus $ A_{{\bf q}\alpha}$
 of wave ${\bf q}\alpha $
has been introduced, 
$\beta_{ikj}$ is the piezo-tensor, ${\mbox{\boldmath $\xi$}}=
{\bf q}/q$, ${\bf q}$ the phonon wave vector,
${\bf e}$ the phonon
unit polarization vector and  $\rho$ is the crystal mass density.
For the crystal of cubic symmetry without an inversion center (class $T_d$)
 tensor $\beta_{ikj}$ has only  non-zero components (all of them equal
 to each other) when all three indexes $i,k,j$ are different, $\beta_{xyz}= \beta_{xzy}=....= h_{14}$.
 For GaAs $eh_{14}= 1.2 \times 10^7 eV/cm$, see, for example, 
 Ref.\onlinecite{Levinson}.
\par
The matrix element for the spin-flip transition between the Zeeman sublevels of orbital
level $n$ with emission of phonon ${\bf q}\alpha$ is:
\begin{equation}
  \left <n \uparrow \mid  U_{ph}^{{\bf q}\alpha}  \mid n \downarrow\right >
 = \sum_{k\ne n}\left[\frac{ (U_{ph}^{{\bf q}\alpha})_{nk}(\hat {\cal H}_1)^{\uparrow  \downarrow}_{kn}}{E_n -E_k - g\mu_B B} +
\frac{ (\hat {\cal H}_1)^{\uparrow  \downarrow}_{nk}
(U_{ph}^{{\bf q}\alpha})_{kn}}{E_n -E_k + g\mu_B B}\right ],
\label{4}
\end{equation}
where   states $n, k$ and corresponding energies $E_n, E_k$ are
determined by first two terms in Hamiltonian (\ref{1}). The spin quantization axis
coincides with the magnetic field vector. In the absence of a magnetic field
the two terms in Eq.(\ref{4}) cancel each other since
$ (\hat {\cal H}_1)^{\uparrow  \downarrow}_{nk} = - (\hat {\cal H}_1)^{\uparrow
\downarrow}_{kn} $ and the matrix
elements involving the phonon operator are symmetric with respect to the interchange
of indexes $n$ and $k$.
This "Van Vleck cancellation" \cite{Van Vleck,Abrahams}  is a consequence of Kramers' theorem and
reduces the matrix element by a factor of
$g\mu_B B/\hbar \omega_0$, $\hbar \omega_0$ being the typical distance between the orbital levels in the dot.
Note that this cancellation occurs for  a spin-orbit Hamiltonian of an arbitrary form.
For instance, it could include the third order terms in the lateral momentum operator. \cite{Khaetskii}
This is in strong contrast with the
cancellation of the linear in the $\beta$ terms in the
matrix elements for the spin-flip
transition between different orbital levels \cite{Khaetskii}, which 
results from the fact that spin-orbit terms $\hat {\cal H}_1$
are linear in the lateral momentum operators, $\hat p_{x,y}$.
Expanding in the above formula with respect to the Zeeman energy, using
  relation $(\hat p_i)_{nk}= (im/\hbar)(E_n-E_k) (x_i)_{nk}$ and
the condition that the
phonon wave length is much larger than the dot size
(i.e. $g\mu_B B \ll \sqrt{ms^2 \hbar \omega_0}$, $s$ is the
sound velocity),
we obtain the efective spin-flip Hamiltonian which acts on the subspace of the
Zeeman sublevels of orbital level $n$:
\begin{equation}
\hat H^{(n)}_{so} = g\mu_B B \frac{m\beta}{\hbar e}
[\hat \sigma_x \alpha_{xx}^{(n)} E_x -\hat \sigma_y \alpha_{yy}^{(n)} E_y +
\frac{(\alpha_{xy}^{(n)}+ \alpha_{yx}^{(n)})}{2}(\hat\sigma_x  E_y -
\hat \sigma_y E_x)],
\label{4.1}
\end{equation}
where $E_{x,y} = - \nabla_{x,y} U_{ph}(x,y)/e$ is the phonon-induced electric field in the
location of the dot.
Here we introduce  polarizability tensor $\hat \alpha$ that may depend on $B_z$. It is given by:
\begin{equation}
 \alpha_{ik}^{(n)}(B_z) = -2e^2\sum_{m\neq n} \frac{(x_i)_{nm}(x_k)_{mn}}{E_n-E_m}
\label{4.2}
\end{equation}
Effective Hamiltonian (\ref{4.1}) is a very general one and can be used to calculate the
spin-flip rates for arbitrary states and dots.
To specify, we will consider only parabolic elliptic dots with the main axes
along the $x,y$ symmetry axes, $\omega_{x,y}$ being the oscillator frequencies. Then the symmetry of
the kinetic
coefficients ensures that $(\alpha_{xy}^{(n)}+ \alpha_{yx}^{(n)}) =0$.
We have to calculate the spin flip matrix elements $<+1/2\mid \hat \sigma_{x,y}
  \mid -1/2>$ over  functions $\Psi_{\mu}, \mu =\pm 1/2 $,
which are the eigenfunctions of operator $\hat\sigma_{z'}$, where the $z'$
axis is directed along the magnetic field vector. These functions are
expressed through the eigenfunctions $\chi_m$ of $\hat\sigma_z$ operator:
$\Psi_{\mu}= \sum_{m=\pm 1/2} D_{\mu m}^{(1/2)\star}
 (\varphi, \vartheta, 0) \chi_m$, where $D^{(1/2)}$ is the finite
rotations matrix \cite{Landau} and   $\varphi, \vartheta  $ are
the azimuthal and  polar angles presenting  ${\bf B}$ in the spherical coordinates.
We substitute $E_{x,y}$ in terms of the boson creation/annihilation operators.
Then for the square modulus of the spin flip matrix element that involves
emission of a phonon with wave vector ${\bf q}$ we obtain:
\begin{eqnarray}
 \mid \hat H^{\uparrow  \downarrow}_{so}({\bf q}\alpha)\mid^2 =
\left (\frac{g\mu_B B m\beta}{\hbar e}\right)^2 A_{{\bf q}\alpha}^2
\left(\frac{\hbar}{2\rho \omega_{{\bf q}\alpha}}\right) \times
\nonumber \\
\{ (\alpha_{xx}^2 q_x^2 + \alpha_{yy}^2 q_y^2)
 \frac{(1+ \cos^2 \vartheta)}{2} - \frac{\sin^2\vartheta}{2}
[(\alpha_{xx}^2 q_x^2 - \alpha_{yy}^2 q_y^2) \cos2\varphi - 2
 \alpha_{xx}\alpha_{yy} q_x q_y \sin2\varphi ] \}
\label{4.3}
\end{eqnarray}
The summing up over all ${\bf q}$ yields the rate due to the first mechanism
\begin{eqnarray}
\Gamma _{1} = \frac{2\pi}{\hbar} \int \frac{d^3 q}{(2\pi)^3} \sum_{\alpha =l,t}
  C_{\alpha} \mid \hat H^{\uparrow  \downarrow}_{so}({\bf q} \alpha)\mid ^2
\delta (\hbar s_{\alpha} q - g\mu_B B) =  \nonumber \\
= \frac{(g\mu_B B)^5}{35\pi \rho \hbar^4}
\left (\frac{h_{14} m\beta}{e \hbar}\right)^2
[(\alpha_{xx}^2  + \alpha_{yy}^2)(1+ \cos^2 \vartheta) -
 (\alpha_{xx}^2 - \alpha_{yy}^2)
\sin^2\vartheta \cos2\varphi]
\left (\frac{1}{s_l^5} + \frac{4}{3s_t^5}\right).
\label{4.4}
\end{eqnarray}
Here $C_l =1, C_t = 2$ and  $s_l, s_t$ are the longitudinal and transverse
 sound velocities. The anisotropy factors  used are:
$A_{{\bf q}, l}^2 = 36 h_{14}^2 \cos^2\theta
\sin^4\theta \sin^2\phi \cos^2\phi$ where $\phi, \theta$ are the azimuthal and
polar angles of vector ${\bf q}$.
$<A_{{\bf q}, t}^2> = 4h_{14}^2 <(\xi_x \xi_y e_z + \xi_x \xi_z e_y+ \xi_y \xi_z e_x)^2> = 2 h_{14}^2 [\cos^2\theta \sin^2\theta +
 \sin^4\theta (1-9\cos^2\theta)\sin^2\phi \cos^2\phi]$, where
 $<...>$ means  averaging over the orientations of the
 ${\bf e}$ vector in the plane which is perpendicular to ${\bf q}$. The
 averaging is done by the formula:
$<e_i e_k> = (1/2)(\delta_{ik} - \xi_i \xi_k)$.
As usual, in the case of finite temperature Eq.(\ref{4.4}) should be
multiplied by  factors $N_{\omega} + 1 \/ (N_{\omega})$ for the
transition with emission (absorbsion) of a phonon, $N_{\omega} =
1/(e^{\hbar\omega/T} -1), \/ \hbar \omega = g\mu_B B$.
Thus, in the case of high temperature $T\gg g\mu_B B$ the
spin-flip rate will be proportional to $(g\mu_B B)^4 T$.
\par
In the particular case of a circular dot $\omega_x = \omega_y =\omega_0$
we have $\alpha_{xx} (B_z) =\alpha_{yy} (B_z) =\alpha_{xx} (0) = e^2/m\omega_0^2 $.
Then, for instance, for the transition between the Zeeman
sublevels of the ground state of the circular dot with emission of a piezo-
phonon we obtain:
 \begin{equation}
\Gamma _{1}
=
\frac{(g\mu_B B)^5}{\hbar(\hbar \omega_0)^4}\Lambda_p (1+ \cos^2 \vartheta); \ \
\Lambda_p \equiv \frac{2}{35\pi}\frac{(eh_{14})^2 \beta^2}{\rho \hbar}
\left (\frac{1}{s_l^5} + \frac{4}{3s_t^5}\right).
\label{6}
\end{equation}
The dimensionless constant $\Lambda_p$ shows the strength of the 
effective spin-piezo-phonon
 coupling in the heterostructure and
ranges from $\approx 7\cdot 10^{-3}$ to $\approx 6\cdot 10^{-2}$ depending on $\beta$.
The spin-flip rate exhibits a very
 strong dependence on the Zeeman energy  and  lateral
confinement energy $\omega_0$. To give a number,
$\Gamma _{1} \approx 1.5\cdot 10^3 s^{-1}$
for $\hbar \omega_0 = 10 K$ and a relatively large magnetic field $B=1T$.
\par
Formula (\ref{4}) is written with allowance for the wave function 
corrections of the 
first order with respect to the spin-orbit Hamiltonian.
The corrections of the second order are described by the 
 following  
 spin-orbit Hamiltonian: 
 \begin{equation}
\hat {\cal H}_{\sigma_z} = 
\frac{m\beta^2}{\hbar} \hat\sigma_z (x \hat p_y -y \hat p_x)
\label{7}
\end{equation}
Then, using this Hamiltonian in formula (\ref{4})  instead of 
$\hat {\cal H}_1$, we can get a nonzero contribution to the spin flip matrix element 
even with zero Zeeman splitting in the denominator (but with taking
into account the orbital magnetic field). 
Keeping again only the term which is linear in ${\bf qr}$  in the 
expansion of  exponent $\exp(i{\bf qr})$, for the rate finally 
we obtain:
     \begin{equation}
\Gamma _{\uparrow,\downarrow}^{(n)}
=\frac{2}{35\pi}
\frac{(g\mu_B B)^3(eh_{14})^2}{\hbar^4 \rho} \left(\frac{m\beta^2}{\hbar}\right)^2 \sin^2\vartheta
\left ( (D_x^{(n)})^2 + (D_y^{(n)})^2\right)
\left (\frac{1}{s_l^5} + \frac{4}{3s_t^5}\right), 
\label{8}
\end{equation}
where 
 \begin{equation}
D_x^{(n)}= 2 Re \sum_{m \neq n}\frac{x_{nm}(\hat L_z + 
(eB_z r^2/2c))_{mn}}{E_n-E_m}, \/ \hat L_z = -i\hbar 
\left (x\frac{\partial}{\partial y}- 
y\frac{\partial}{\partial x}\right).
\label{8.1}
\end{equation}
In the absence of the magnetic field quantities $D_x,\/D_y$ are 
identically equal to zero. 
Using the properties of the matrix 
elements for the linear oscillators
we obtain 
that in the case of elliptic (circular) dots $D_x=D_y=0$. 
Keeping the term which is quadratic in ${\bf qr}$  in the 
expansion of the exponent $\exp(i{\bf qr})$, we  obtain 
 a non-zero contribution but the corresponding spin-flip
rate  is smaller than contribution $\Gamma_1$ by a
factor of $(\beta/s)^2 
(\omega_c/\omega_0)^2 < 1$, here
$\omega_c = eB_z/mc$. It is also clear  that in the case of irregular
dots  quantities $D_{x,y}$ are not equal to zero. The ratio of the 
corresponding rate and $\Gamma_1$ can be 
estimated as 
$\tau^2 (m_0 \beta a/\hbar )^2$, where $\tau$ is a dimensionless 
parameter which describes the deviation from ellipticity and $a$ is a dot size.
Even when $\tau  \simeq 1$ 
 this ratio is of the order of unity 
 for a typical dot size $a \approx 10^3 \AA $.  
 Therefore, for $\tau \ll 1 $ we can expect that contribution 
Eq.(\ref{8}) is much
smaller than $\Gamma_1$.
\par Note that, besides  term $ \hat {\cal H}_1 $ which is 
linear in the 2D momentum,  the initial 
Hamiltonian  also contains the
term which is cubic in the momentum:
$(1/2)\hat \sigma_x \{\hat p_x, \hat p_y^2 \}_+ - 
(1/2)\hat \sigma_y \{\hat p_y, \hat p_x^2 \}_+$.
Again, in the presence of the orbital magnetic field we could get some
contribution to the spin-flip rate. 
To this end, we need to calculate  quantities 
$\tilde D_x,\/\tilde D_y$  obtained from 
$D_x,\/ D_y$ by replacing  operator 
$ \hat L_z + 
(eB_z r^2/2c)$ by $(1/2)\{\hat p_x, \hat p_y^2 \}_+$
 or $(1/2)\{\hat p_y, \hat p_x^2 \}_+$.
  In the case of elliptic (circular) dots
we obtain 
$\tilde D_x=\tilde D_y=0$ because of the symmetry.

    \subsection{Direct spin-phonon coupling.}

Using the standard presentation
for the strain tensor in terms of the acoustic
phonon modes, we calculate the matrix element of  $\hat {\cal H}_2$
for the  electron spin-flip transition
between the Zeeman sublevels of  orbital state $\Phi$ with emission
of a phonon with momentum ${\bf q}$:
\begin{equation}
M_{\uparrow,\downarrow} =  \frac{V_0}{4}
\left (\frac{\hbar}{2\rho \omega_q} \right )^{1/2}[q_x e_y + q_y e_x] \left <\Phi\mid \frac{1}{2}
\{ (\hat p_x +i \hat p_y), \exp(i{\bf q r})\}_+ \mid \Phi \right >.
\label{2}
\end{equation}
For simplicity, here we set  ${\bf B} \parallel z$. Similar expressions
were obtained in Ref. \onlinecite{Frenkel} for a different problem.
The total spin-flip rate is given by the Fermi golden rule:
\begin{equation}
\Gamma _2 = \frac{\pi \hbar V_0^2}{16 \rho g\mu_B B}  \int \frac{d^3 q}{(2\pi)^3}
 (q_x^2 + q_y^2) \mid  \left <\Phi \mid \frac{1}{2}
\{ (\hat p_x +i \hat p_y), \exp(i{\bf q r})\}_+ \mid \Phi \right > \mid ^2  \delta (\hbar s q - g\mu_B B).
\label{3}
\end{equation}
The relevant phonon wave length
is much larger than the dot size, which allows for further simplifications.
We concentrate on a
circular dot with confining frequency $\omega_0$.
For the orbital states
with $n=0$ and  $l=0,\pm 1$ (the ground  and the first two excited states):
 \begin{equation}
\Gamma _2 =
\frac{V_0^2 (g\mu_B B)^5}{240 \pi \rho s^7 \hbar^4} [l+ \frac{\omega_c}{2 \sqrt{\omega_0^2 + (\omega_c^2/4)}} (\mid l\mid + 1)!]^2.
\label{5}
\end{equation}
The spin-flip rate produced by term  $\hat {\cal H}_3$   does not depend on the
structure
of the orbital state and is given by
\begin{equation}
\Gamma_3  \simeq \frac{(\mu_B \tilde g B)^2(\mu_B g B)^3}{\rho s^5_t \hbar^4}
\end{equation}
Let us now compare the rates $\Gamma_{1,2,3}$ obtained.
All of them are proportional to the fifth power of  energy splitting $g \mu_B B$,
so that their ratio hardly depends on the magnetic field.
 First, we note that the ratio of $\Gamma_3$ and $\Gamma_2$ is of the order
of $(\tilde g s_t/g V_0)^2 \approx 7.8 \cdot 10^{-3}\ll 1$. So that $\Gamma_2$ is more important.
The ratio of $\Gamma_1$ and $\Gamma_2$ is of the order of
$ (eh_{14}/mV_0/\hbar)^2 (m\beta^2 ms^2_t/(\hbar \omega_0)^4)$.
For $ \hbar \omega_0 \simeq 1 \div 10 K $  the ratio is of
 the order of $10^{6} \div 10^{2}$ and  increases only for larger dots that have
 smaller $\omega_0$.
Thus, we conclude that the admixture mechanism dominates.

 \subsection{Two-phonon processes.}

     The calculated rate $\Gamma_1$ is
small partly because of the small phonon  density of the 
states at the scale of
the Zeeman energy. 
On the other hand, for the case of the spin-flip
transitions between the Zeeman levels of usual impurity \cite{Abrahams}
 the two phonon  processes under some conditions
may become more important than the single phonon processes. 
At sufficiently small Zeeman splitting 
the contribution of the single phonon processes is very small, and 
with increasing temperature the role of the processes when one phonon 
is absorbed and the other is emitted is increased.
It is also true for the case of a quantum dot and here we 
  give some formulas which describe the 
contribution of such two phonon  processes for GaAs quantum dots 
  in several limiting cases.
We also indicate the conditions under which these contributions 
can be important. 
\par
If we treat the interaction with the phonons in 
the second order, we obtain processes in which a phonon is scattered from
 state ${\bf p}$ to  state ${\bf q}$ while the electron spin flips. 
The effective matrix element contains transitions to an excited orbital state
 with the emission or absorption of a phonon and then transitions back to
 the ground state with the absorption or emission  of a phonon.
 The spin may flip either in the first or second  transition. 
 The matrix element is \cite{Abrahams}:
 \begin{equation}
 <V_2> \sim  \left(\frac{\hbar}{\rho s \sqrt{pq}}\right) 
 (eh_{14})^2 [N_p (N_q + 1)]^{1/2} \sum_a \left \{ 
 \frac{[H_{p,-q}^+ + H_{p,-q}^-]}{-\Delta_a - \hbar s q}
 +  \frac{[H_{-q,p}^+ + H_{-q,p}^-]}{-\Delta_a + \hbar s p}
  \right \}, 
 \label{two}
 \end{equation}
 $$ H_{p,q}^{\pm}= (\Psi_0^+, \exp(i{\bf pr})\Psi_a^{\pm})
(\Psi_a^{\pm}, \exp(i{\bf qr})\Psi_0^{-}), $$
 where $N_p$ is the Bose distribution function and $\Delta_a$ is the energy 
separation between the ground state whose  wave function is  $\Psi_0$
 and the excited state whose wave function for spin up, say, is 
$\Psi_a^+$. We can neglect the Zeeman energy in the denominators 
since no Van Vleck cancellation  occurs here. 
We consider  again interaction with piezo-phonons 
since deformation phonons become important at very high temperature 
(see below). 
For simplicity we consider here only the case when the magnetic field 
is perpendicular to the 2D plane and study the relaxation of $S_z$ 
spin component.  
 \par        
 As it was shown in Ref.\onlinecite{Khaetskii}, 
there is a cancellation of the linear in the $\beta$ terms in the
matrix elements of  type 
$(\Psi_0^+, \exp(i{\bf pr})\Psi_a^{-})$ 
 for the spin-flip
transition between different orbital levels.
 This is a consequence
of the fact that spin-orbit terms $\hat {\cal H}_1$
are linear in the lateral momentum operators, $\hat p_{x,y}$.
For that reason, quantities  $ H_{p,q}^{\pm}$ 
are  proportional to the first power of $\beta$ only
if one takes into account  the Zeeman splitting in the 
electron spectrum.\cite{Khaetskii} 
We consider the temperature in the interval $g\mu_B B \ll T \ll \hbar \omega_0$,
where $\hbar \omega_0$  is the characteristic energy distance between the levels
in the dot. Since $\hbar s p \simeq \hbar s q < T$, 
then we can neglect the phonon energies in the denominators while 
calculating the contribution  to $<V_2>$ proportional to the first power of
 $\beta$.
 It is apparent  that the spin-flip rate has a different temperature 
 dependence for the  temperatures smaller and larger than $T_0 \simeq 
 \sqrt{ms^2 \hbar \omega_0}$. At this characteristic temperature 
 the phonon wave length is equal to  lateral dot size $\lambda$.
 For GaAs at  $ \hbar \omega_0 \simeq 10 K$ temperature $T_0 \approx 1 K$. 
 Let us give the estimate for the spin-flip rate in the case 
 $T\ll T_0$, when $p\lambda, q\lambda \ll 1$.
  Here the estimate for $ H_{p,q}^{\pm}$ is  
$ H_{p,q}^{\pm} \simeq (\beta/\lambda \omega_0 ) (g\mu_B B/ \hbar \omega_0) 
(\lambda q)^3$. Then the relaxation rate is:
\begin{equation}
 \Gamma^{(2)}_1 (T) = (2\pi/\hbar) 
       \sum_{p,q} \mid <V_2> \mid^2 \delta [\hbar s (p-q) - g\mu_B B] 
       \simeq \frac{\Lambda_p^2}{\hbar} 
       \frac{s^2}{\beta^2}
        \frac{(g\mu_B B )^2 (ms^2)^{5/2}}{(\hbar \omega_0)^{7/2}}
    \left (\frac{T}{T_0} \right)^9
\label{two1}
 \end{equation}
In performing the integral over ${\bf p}$, we have neglected $ g\mu_B B$ 
in comparison to $\hbar s p $. 

\par 
In the case  $T \gg T_0$ 
 the  momentum components parallel to
the plane are estimated as $q_{\parallel} \lambda \simeq 1$, otherwise 
the matrix elements are exponentially small.
As to the $q_z$ values, the contribution of the region 
$q_z \simeq T/\hbar s \gg q_{\parallel}$ is much smaller than 
that where  $q_z \simeq q_{\parallel} \ll T/\hbar s$. 
Such is the case  even without regard for
the fact that for the [100] orientation of the 2D plane
 the effective piezo-modulus 
    $ A_{{\bf q}\alpha}$ introduced above has a smallness 
    $q_{\parallel}/q_z \ll 1$. 
  Thus, calculating the contribution from  
  $q_z \simeq q_{\parallel} \ll T/\hbar s$ and taking into account 
  that $N_p = T/\hbar s p \gg 1$,  we obtain
     for the spin-flip rate in the case   $  T_0 \ll T \ll \hbar \omega_0$:
\begin{equation}
 \Gamma^{(2)}_2 (T)
       \simeq \frac{\Lambda_p^2}{\hbar} 
       \frac{s^2}{\beta^2}
        \frac{(g\mu_B B )^2 (ms^2)^{5/2}}{(\hbar \omega_0)^{7/2}}
    \left (\frac{T}{T_0} \right)^2
\label{two2}
 \end{equation}
  The contribution from the deformation phonons is much smaller.
  In the case $T\gg T_0$ the characteristic  $q_z \simeq T/\hbar s \gg 
 q_{\parallel}\simeq 1/\lambda$, i.e. the  deformation phonons
 are emitted almost perpendicular to the 2D plane. 
 Then, for the deformation potential contribution to the spin-flip rate 
 we obtain:
 $\Gamma_d \simeq (\Lambda_d^2 /\hbar)(\beta^2/s^2)
  (g\mu_B B )^2 T^3/(\hbar \omega_0)^3
ms^2 $, where $\Lambda_d \simeq (1/2\pi) (\Sigma^2 m^2/\rho \hbar^3 s)$
is the  dimensionless constant  which shows the strength of the 
electron interaction with  deformation  phonons. 
 For GaAs  $\Lambda_d \approx  10^{-5}$.  
 Even at $T \simeq \hbar \omega_0 $
   the value of 
   $\Gamma_d /\Gamma^{(2)}_2  \simeq (\hbar \omega_0/ms^2)^{5/2} 
   (\Lambda_d \beta^2/\Lambda_p s^2)^2$ is much  smaller than unity 
   for any realistic $\hbar \omega_0$. For example,
   at  $\hbar \omega_0 =30 K$ this ratio is $\approx 0.03$.
\par
  Let us compare the 
 two-phonon contribution $\Gamma^{(2)}_2 $
 with $\Gamma_1 T/g\mu_B B $. 
  We see that the  two-phonon contribution $\Gamma_2^{(2)}$ 
 prevails at sufficiently small Zeeman splittings:
 $g\mu_B B < ms^2 (\Lambda_p s^2/\beta^2)^{1/2} (T/T_0)^{1/2}$.
Taking  the maximal temperature $ T \simeq \hbar \omega_0 $, 
we obtain  $g\mu_B B < [(\Lambda_p s^2/\beta^2)
 \sqrt{ms^2/\hbar \omega_0}]^{1/2} T_0$.  For
$\hbar \omega_0 \approx 10 K$ we see that
this contribution is more important for magnetic fields  smaller 
than approximately 0.4 T (where the estimate for the spin-flip 
time is of the order of ms).   
On the other hand, at  $T \simeq T_0$
we obtain  for the same $\hbar \omega_0$ that $g\mu_B B <
0.03 K$ (i.e. the two-phonon contribution is more important for magnetic fields 
smaller than $\approx 1 kG$). 
For these fields 
the characteristic spin-flip time is of the order of 1s, 
i.e. it is still long.   
\par The general conclusion is that 
at sufficiently low temperatures 
(much smaller than $\hbar \omega_0$) the characteristic Zeeman splittings
below which the two-phonon contribution to the spin-flip rate dominates
are small and corresponding spin-flip times are unusually long 
(see the estimates above).

 \section{Other mechanisms of the spin flip.}

Let us discuss briefly  other mechanisms of the  spin flip.
The spin transitions between the Zeeman sublevels of the impurity   state
in semiconductors ( mostly Si) were extensively studied quite a long
time ago. \cite{Pines,Abrahams}
Except spin-orbit coupling, several other mechanisms were proposed, such as:
1) modulation of the hyperfine coupling with nuclei by lattice vibrations,
2) the spin-spin interaction between the bound electron and the conduction
electron in the leads,
3) the spin-current interaction, when the bound electron spin flip is caused
by the fluctuating magnetic field of the conduction electrons,
4) an exchange scattering process which flips the spins of
both the conduction electron and the bound donor electron.
 Whereas  the spin-orbit interaction strongly depends on
the  crystal symmetry and is different  for Si and GaAs,
the other
mechanisms are quite general in  nature and we can profit
 from the discussion in \cite{Pines,Abrahams}.

\par
Mechanism (4) requires an overlap of the wave functions of the electrons in
the leads and in the dot. In the context of QD it is considered in Ref. \onlinecite{Glazman}.
The corresponding rates are not intrinsic to the dot since they are
proportional to the barrier transparencies. They can be tuned to arbitrary low values.
Refs.(\onlinecite{Abrahams,Pines}) have demonstrated that the spin-flip rates
associated with mechanisms (2,3) are very small.
As an example, we give the rate estimation for  mechanism (3).
The Bio-Savar formula relates the magnetic field and current fluctuations
in the leads so that
$<H^2>_{\omega} \simeq (1/c^2 a^2) <I^2>_{\omega}$,
$a$ being the characteristic distance between
the electron in the dot and the electrons in the leads. Using the Nyquist formula
for the  correlator of the currents we estimate
$<H^2>_{\omega} \simeq (1/c^2 a^2) \hbar\omega \coth (\hbar \omega /2T) (1/R)$,
$R$ being the typical resistance of the leads or the dot environment.
Thus, the corresponding   spin-flip relaxation rate is  estimated as
\begin{equation}
\Gamma_4 \simeq \mu^2_B <H^2>_{\omega} /\hbar^2
 \simeq \omega  \big(\frac{\lambda_c}{a}\big)^2
\frac{\hbar}{e^2 R}
\end{equation}
where $\lambda_c = e^2/m_0 c^2 \approx 2.8\cdot 10^{-13} $ cm
is the classical electron radius, $\hbar \omega= g\mu_B B$.
Rate $\Gamma_4$ is proportional to the first power of  Zeeman splitting
so that it may formally compete with $\Gamma_1$ at sufficiently small splittings.
However, this occurs at splittings that are so small that the corresponding rates
are not observable. To give an example, we choose $R=1 Ohm$ and
$\hbar \omega_0 = 1 K$, which corresponds to $a \simeq 1.15\cdot 10^{-5} cm$.
Then  rate $\Gamma_4$
would dominate if splitting $\hbar \omega \ll 2.5\cdot 10^{-3}K$. This
corresponds to the rates lower
than $8\cdot10^{-4} s^{-1}$!
\par As to mechanism (1), 
i.e. modulation of the hyperfine coupling with nuclei by lattice vibrations,
the relative strength of this mechanism and the spin-orbit interaction 
 can be different for different 
materials. For example, in the case of Si where the 
spin-orbit interaction   is much weaker than in GaAs, the 
dominant mechanism of the spin-flip for the case of the phonon assisted 
transitions between the Zeeman levels of usual impurity (the situation studied
in Ref. \cite{Pines}) was found to be the 
modulation of the hyperfine coupling with nuclei by lattice vibrations.
In the case of GaAs, however, our conclusion is that 
the dominant mechanism is the admixture mechanism of the 
spin-orbit interaction.
This conclusion was reached for the first time in Ref.\onlinecite{Me}, 
where the calculations used  essentially followed those in 
Ref.\onlinecite{Pines}.
Here we give  the result obtained in Ref.\onlinecite{Me} for the rate 
 due to  the modulation of the hyperfine coupling with nuclei 
 by lattice vibrations
\begin{equation}
\Gamma_h \simeq (g\mu_B B)^3 \gamma^2 \omega_N^2 /\hbar^2
s^5 \rho,
\label{hyper}
\end{equation}
where $\omega_N \simeq (v_0A^2/a^2 z_0)^{1/2}/\hbar$ is
the electron spin precession frequency in the random 
field of unpolarized nuclei, $v_0$ is the unit cell volume, $A$ the
hyperfine  interaction constant, $a$ the dot lateral  size and $z_0$ the electron wave
function extension in the z-direction. 
Finally,  $\gamma \simeq (1/m) (dm/d\Delta)$ is 
the change in  effective mass $m$ with dilation, see also 
Ref. \onlinecite{Pines}. 
For GaAs QD with $a \approx 10^3 \AA$ and $z_0  \approx 10^2 \AA $,
$\omega_N$ can reach $\simeq 10^8 s^{-1}$.  
  Let us compare  spin-flip rate Eq.(\ref{hyper})
   with $\Gamma_1$ (this comparison was done earlier in Ref.\onlinecite{Me}).
Even taken for $\gamma \simeq 50$ (see also \cite{Pines}) we can easily
see that $\Gamma_h$ will
compete with $\Gamma_1$ at the Zeeman  splitting 
$\simeq \gamma \omega_N (\hbar \omega_0)^2/ \beta eh_{14}$ 
which is  so small that the corresponding rate
is not observable. For example,  for $\hbar \omega_0 = 10 K$  the splitting is 
of the order of $10^{-5}$ K. 
 Therefore,   the admixture mechanism 
 of the spin-orbit  interaction is the dominant one. 
\par 
It should be noted, that 
in this work we have not considered  the electron spin relaxation mechanism which is through the hyperfine interaction  related to the internal nuclear dynamics. The latter is due to the dipole-dipole interaction between the nuclei which does not conserve the total nuclear spin \cite{Abragam}.  This mechanism might be important at low magnetic fields.
However, this problem is not simple and needs a seperate investigation.

\par
Finally, we mention the experimental studies of spin relaxation 
in n-type GaAs quantum dots. 
Such an experiment  
has been recently carried out \cite{Fujisawa}.
The non-equilibrium tunneling 
current through excited states in an AlGaAs/GaAs quantum dot was studied 
using a pulse-excitation technique which measures the energy relaxation time from 
the excited state to the ground state. 
Some excited states showed a relaxation time which was much longer than a 
few $\mu s$, while the other showed time much shorter than a few ns. 
This great difference in relaxation times was ascribed to the fact that 
some inelastic transitions are accompanied  by the spin flip. 
For these transitions the relaxation time was so long that the
method used in the above mentioned paper only allowed to give some
estimation (much longer than a few $\mu s$).  Though the 
transitions studied by T.Fujisawa et al. could    in general
involve the spin flip transitions between the states with {\it different }
orbital structures (this situation was considered in our previous paper 
\cite{Khaetskii}), 
the experimental data confirm the general statement that the spin-flip 
processes in n-type quantum dots can be really slow. 
 \par
In conclusion, we have calculated the rates for the phonon-assisted
spin-flip transitions  between the Zeeman sublevels in a quantum dot for
all possible
mechanisms and  shown that the admixture mechanism of the 
spin-orbit interaction is a dominant one.
 The corresponding spin-flip rate $\Gamma_1$ (see Eqs.(\ref{4.4},\ref{6}))
 exhibits a strong dependence on Zeeman energy and at small magnetic fields
takes very low values (up to seconds).

This work is  part of the research program of the "Stichting voor Fundamenteel
Onderzoek der Materie (FOM)". We acknowledge
the support of the Netherlands Organization for Scientific Research (NWO)
in the framework of Dutch-Russian collaboration and the NEDO project NTDP-98.
We are grateful to  L.P. Kouwenhoven, T.H. Oosterkamp, G.E.W. Bauer,
 T. Fujisawa, Y. Tokura, Y. Hirayama and D. Loss for useful discussions.

\end{document}